\documentclass{iaus}
\usepackage{graphicx}

\title[SK\,1: A Possible Case of Triggered Star Formation in Perseus] 
{SK\,1: A Possible Case of Triggered Star Formation in Perseus}

\author[Rengel, Hodapp and Eisl\"offel]   
{Miriam Rengel$^1$
, Klaus Hodapp$^2$ \break \and Jochen Eisl\"offel$^3$}

\affiliation{$^1$Max Planck Institute for Solar System Research,
Katlenburg-Lindau, 37191, Germany \break email: rengel@mps.mpg.de\\[\affilskip]
$^2$Institute for Astronomy, 640 N. A'hookup Place, Hilo, HI
96720 \break email: hodapp@ifa.hawaii.edu\\[\affilskip]
$^3$Th\"uringer Landessternwarte Tautenburg,  Sternwarte 5, 07778
Tautenburg, Germany \break email: jochen@tls-tautenburg.de}

\pubyear{2006}
\volume{237}  
\pagerange{119--126}
\date{?? and in revised form ??}
 \jname{Triggered Star Formation in a
Turbulent ISM} \editors{B. G. Elmegreen \& J. Palous, eds.}
\begin{document}

\maketitle

\begin{abstract}
According to a triggered star formation scenario (e.g.
\cite[Martin-Pintado \& Cernicharo 1987]{mc87}) outflows powered by
young stellar objects shape the molecular clouds, can dig cavities,
and trigger new star formation. NGC\,1333 is an active site of low-
and intermediate star formation in Perseus and is a suggested site
of self-regulated star formation (\cite[Norman \& Silk 1980]{ns80}).
Therefore it is a suitable target for a study of triggered star
formation (e.g. \cite[Sandell \& Knee 2001]{sk01}, SK\,1).  On the
other hand, continuum sub-mm observations of star forming regions
can detect dust thermal emission of embedded sources (which drive
outflows), and further detailed structures.

Within the framework of our wide-field mapping  of star formation
regions in the Perseus and Orion molecular clouds using SCUBA  at
850 and 450 $\mu$m, we map NCG\,1333 with an area of around 14$'$
$\times$ 21$'$. The maps show more structure than the previous maps
of the region observed in sub-mm. We have unveiled the known
embedded SK\,1 source (in the dust shell of the SSV\,13 ridge) and
detailed structure of the region, among some other young protostars.

In agreement with the SK\,1 observations, our map of the region
shows lumpy filaments and shells/cavities that seem to be created by
outflows.  The measured mass of SK\,1 ($\sim$0.07 M$_{\bigodot}$) is
much less than its virial mass ($\sim$0.2-1 M$_{\bigodot}$). Our
observations support the idea of SK\,1 as an event triggered by
outflow-driven shells in NGC\,1333 (induced by an increase in gas
pressure and density due to radiation pressure from the stellar
winds, that have presumably created the dust shell). This kind of
evidences provides a more thorough understanding of the star
formation regulation processes. 
\end{abstract}

\firstsection 
\section{Introduction}

When an area of active or recent star formation is found, a question
can be raised: is the star formation there happening spontaneously
or is it being triggered? Places where there is no apparent trigger
would be a possible evidence of its spontaneous nature, but in areas
where there are indications of triggered star formation the
determination of the possible causes is not easy. Star forming areas
contain population of young stellar objects (YSOs), then the
mentioned question can be addressed studying the interactions
between components of the YSOs and their surrounding Inter Stellar
Medium (ISM).

Powerful outflows are associated with strong accretion activity of
the earliest protostellar phase (Class\,0 stage) in the formation of
a low-mass star. The Class\,0 phase was identified by \cite[Andr\'e;
Ward-Thompson \& Barsony 1993]{awb93}, and consists of a central
protostellar object, surrounded by an infalling envelope, and a
flattened accretion disk. The later SED classes (1,2 and 3, e.g.
\cite[Lada 1987]{l87}) are characterized by progressively
diminishing accretion rates and consequently, less outflow power
(\cite[Bontemps; Andr\'e; Terebey; et al. 1996]{batc96};
\cite[Henriksen; Andr\'e  \& Bontemps 1997]{hap97}; \cite[Davis \&
Eisl\"offel 1995]{de95}).

The Class\,0  phase is of special interest, not only because most of
the characteristics of a future star are determined during this
phase, but outflows can have a profound effect on the surrounded
ISM: they shape the molecular clouds, can dig cavities, compress
dust shells and trigger new star formation. Observations of this
phase are, however, difficult for two reasons: first, the hot,
near-stellar core of a Class\,0 object is so heavily obscured (A$_V$
$\approx$ 500 mag) as to make the object undetectable up into the
mid-IR. Second, the Class\,0 phase is of short duration; presumably
of order of a few $10^5$yr (\cite[Visser; Richer \& Chandler
2002]{vrc02}). Therefore, only a small number of these objects has
been found to allow detailed studies. Nevertheless, continuum sub-mm
observations of these sources can detect dust thermal emission of
the circumstellar envelopes and provide a powerful tool for
constraining the distribution of matter in Class\,0 objects
(\cite[Adams 1991]{a91}).

As part of a more extensive study of star-forming regions and of the
physical structure and processes in Class\,0 sources (\cite[Rengel
2004]{r04}), we report here the possibility of triggered star
formation by outflows driven by YSOs and discuss evidence for this
triggering.

\section{Imaging of the molecular cloud NGC\,1333 and results}\label{sec:imaging}

\subsection{Target Selection}
Together with the regions L1448, L1455, HH211 (in Perseus) and L1634
and L1641 N (in Orion), we map NCG\,1333 at 850 and 450 $\mu$m using
the Submillimetre Common User Bolometer Array (SCUBA) camera at the
James Clerk Maxwell Telescope (JCMT). NGC\,1333 is an active site of
low- and intermediate star formation in Perseus and is a suggested
site of self-regulated star formation (\cite[Norman \& Silk
1980]{ns80}). It has been observed in the sub-mm by \cite[Looney;
Mundy \& Welch 2000]{lmw00}, \cite[Sandell \& Knee 2001]{sk01}, and
\cite[Chini; Ward-Thompson; Kirk; et al. 2001]{cwkrs01}.
NGC\,1333\,IRAS\,2 (\cite[Jennings; Cameron; Cudlip; et al.
1987]{jcch87}) is located at the edge of the large cavity
(\cite[Langer; Castets \& Lefloch 1996]{lcl96}). It has been
resolved into three sources: 2A and 2B, detected by several authors
(\cite[Sandell; Knee; Aspin; et al. 1994]{skarr94}; \cite[Blake;
Sandell; van Dishoeck; et al. 1995]{bsv95}; \cite[Lefloch; Castets;
Cernicharo; et al. 1998]{lccl98}; \cite[Rodr\'{i}guez; Anglada \&
Curiel 1999]{1999}; \cite[Looney; Mundy \& Welch 2000]{lmw00};
\cite[Sandell \& Knee 2001]{sk01}; \cite[J$\o$rgensen; Hogerheijde;
van Dishoeck; et al. 2004]{jhv04}), and 2C, detected by
\cite[Sandell \& Knee 2001]{k01}. IRAS\,2 A and B are Class\,0
candidates (\cite[Sandell \& Knee 2001]{sk01}, and \cite[Motte \&
Andr\'e 2001]{ma01}). Observations of IRAS\,2 (A,C) in mid-IR are
reported by \cite[Rebull; Cole; Stapelfeltd; et al. 2003]{rcsw03}.

\subsection{Observations}
NGC\,1333 was mapped here with an area of 14$'$ $\times$ 21$'$ with
both jiggle and scan maps. Data reduction treatment is described in
\cite[Rengel 2004]{r04}. These new maps show more structure than
previous sub-mm maps of this region. We include SSV 13, south areas,
the region surrounded IRAS\,1 and NGC\,1333\,S. This later region
was first noted by \cite[Rengel; Froebrich; Hodapp; et al.
2002]{rfhe02}, further discussed by \cite[Rengel 2004]{r04} and
independently discovered by \cite[Young; Shirley; Evans; et al.
2003]{y03}. \cite[Hodapp; Bally; Eisl\"offel; et al. 2005]{hbed05}
discussed a subset of the sub-mm data presented here in their
relation to NIR observations and pointed out that the driving source
of the system HH343A-F--HH340B was likely associated with the
easternmost clump in NGC\,1333\,S. Newly identified structure
introduced by \cite[Rengel; Hodapp \& Eisl\"offel 2005]{r05}
strengthen the tentative conclusion reached by \cite[Hodapp; Bally;
Eisl\"offel; et al. 2005]{hbed05} that NGC\,1333\,S is the site of
secondary, low-mass star formation triggered by the powerful IRAS
1-9 protostars about 1 pc of this region.

Furthermore, SK\,1 is an embedded source in the dust shell south of
the SVS\,13 ridge, with an emission (east) detected in the 850
$\mu$m map. We unveiled it, and found several dust ridges and shells
formed by outflows in the region, among further detailed structure
and some other YSOs (Fig.\,1). \cite[Sandell \& Knee 2001]{sk01}
suggest that SK\,1 appears to be a case of triggered star formation
by outflow-driven shells. Further details of the mapped region are
given in \cite[Rengel 2004]{r04} and \cite[Rengel, Hodapp \&
Eisl\"offel 2006]{rhe06}.

\begin{figure}
\includegraphics[angle=-90,width=12cm]{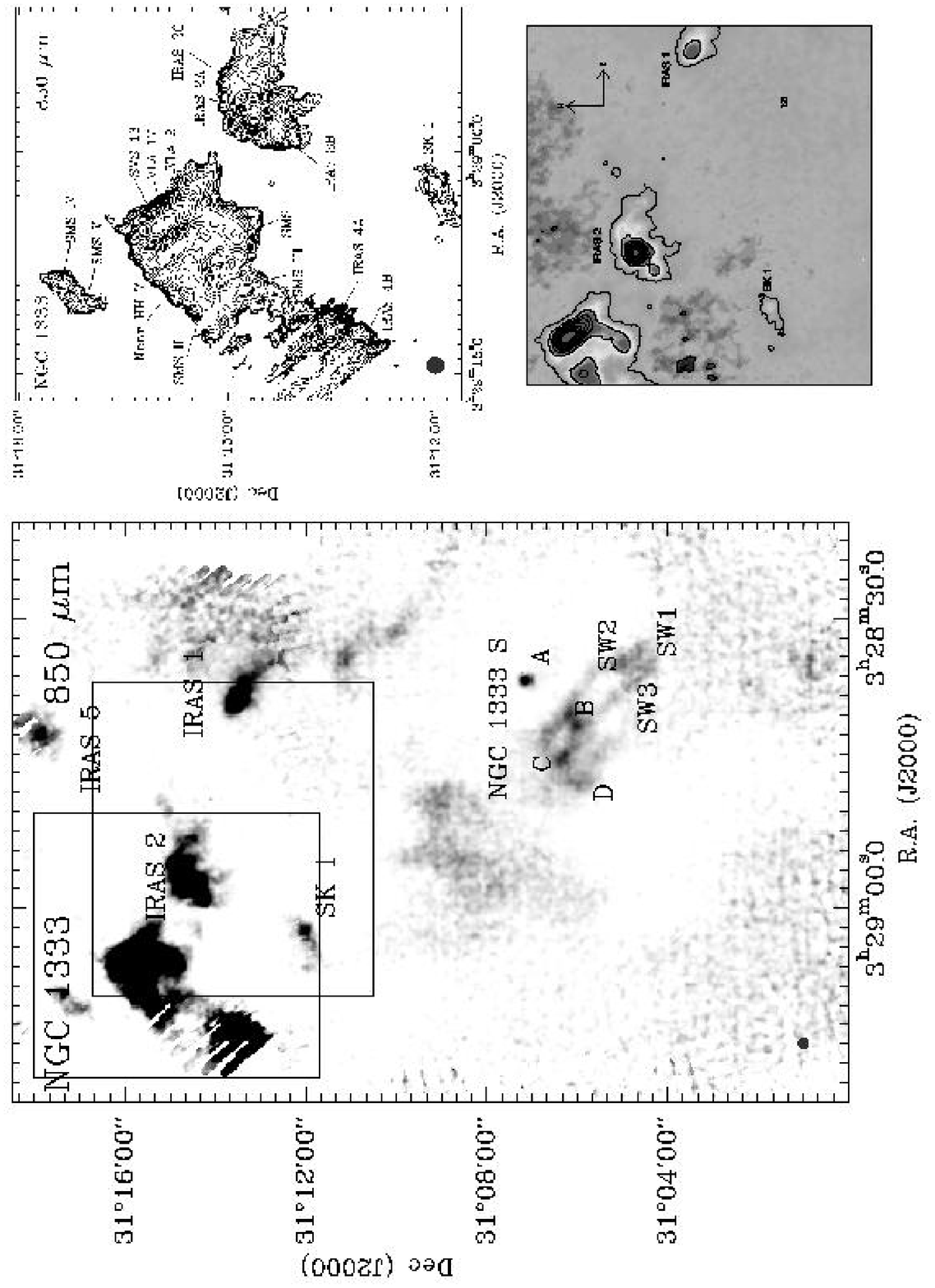}
\vspace{-0.1cm}
 \caption{\label{ssv13}
{\small The 850 $\mu$m deep field map of NGC\,1333 (left) and the
850 $\mu$m contour map of the SSV\,13 area in NGC\,1333 (upper
right) (left box in left figure). Contour levels are in log scale
with step size 1.25 from 0.0002 Jy\,beam$^{-1}$. The filled circles
in the bottom left indicate the main beam size. Down right image
shows the deep field map of the region containing SK\,1.}}
\end{figure}

\section{Deriving physical parameters}

We estimate the gas and dust masses for the sample from the dust
emission according to the mass equation of \cite[Hildebrand
1983]{h83}. We find a mass of 3.7 M$_{\bigodot}$ for IRAS\,2A and
$\sim$0.07 M$_{\bigodot}$ for SK\,1.

We infer the bolometric temperature and luminosity of IRAS\,2A and
SK\,1 by constructing the complete SEDs (we combine the SCUBA fluxes
with data from the literature in several wavelengths (\cite[Rengel;
Froebrich; Wolf; et al. 2004]{rfwe04})). Ages are further calculated
following the Smith protostellar evolutionary scheme (\cite[Smith
1999]{s99}, \cite[Smith 2002]{s02}). For IRAS\,2A and SK\,1, we find
ages of 26 and 15 $\times$ 10$^{3}$ yr, respectively.

Within the framework of deriving physical parameters of the YSOs
contained our wide-field mapping of star formation regions, we also
establish further physical conditions (e.g. temperature
distribution, radius, and power-law index of the density) of some
sources in the sample, including IRAS\,2A. This is performed by
using the radiative transfer code MC3D (\cite[Wolf; Henning \&
Stecklum 1999]{w99}). Further modeling details are given in
\cite[Rengel 2004]{r04} and \cite[Rengel, Hodapp \& Eisl\"offel
2006]{rhe06}.

\section{SK\,1: Suspected triggered star formation in NGC\,1333}\label{sec:tsf}

Are the molecular outflows of IRAS\,2 (\cite[Knee \& Sandell
2000]{ks00}; \cite[J${\o}$rgensen, Hogerheijde, Blake, et al.
2004]{jhb04}) perhaps the main cause of the formation of SK\,1? We
report here some evidences that support the idea of modification and
disruption of areas by the IRAS\,2 outflows, and the formation of
SK\,1 as result of the compressed dust shells/cavities.

First, our map of the region shows lumpy filaments and
shells/cavities that seem to be created by outflows. Second, the
measured mass of SK\,1 is much less than its virial mass
($\sim$0.2-1 M$_{\bigodot}$). This indicates that SK\,1 is not in a
stationary state, isolated and self-gravitating bound. Third,
regarding the star formation timing, a hard upper limit is placed on
the age: IRAS\,2A is older than SK\,1. Fourth: which one physical
mechanism is the responsible of the formation of SK\,1? we rule out
cluster dissipation as the main mechanism: if a standard cluster
dispersion velocity of 1 km s$^{-1}$ is assumed, SK\,1 has not
traveled far. ``Collect-and-collapse" (CAC) and radiation-driven
implosion (RDI) scenarios (\cite[Elmegreen \& Lada 1977]{el77},
\cite[Oort \& Spitzer 1955]{os55}) have been proposed as physical
mechanisms for the star formation at the edge of an HII region.
Timescales for the RDI mechanism are less than for the CAC
scenarios, then it seems that this later one is the most supported.
Fifth, IRAS\,1 has a long tail pointing away from the HII region,
which is an indication of direct interaction between the outflows
and the gas and dust in the vicinity.

\section{Conclusion}\label{sec:conclu}

SK\,1 provides an interesting study of interactions between outflows
and surrounding ISM. In at least one case, we identify this
protostellar source whose formation is likely to have been triggered
by powerful outflow bow shocks. This kind of evidences provides a
more thorough understanding of the star formation regulation
processes. Further observations of molecular gas are necessary.

\begin{acknowledgments}
We thank the S237 organizers, and T. Jenness for the assistance with
SURF.
\end{acknowledgments}

\end{document}